\documentclass[letterpaper, 10pt, conference]{ieeeconf}

\IEEEoverridecommandlockouts                              
\usepackage{graphicx}

\newtheorem{theorem}{Theorem}[section]
\newtheorem{lemma}[theorem]{Lemma}

\newcommand{\sr}{\stackrel}

\newcommand{\tri}{\sr{\triangle}{=}}

\newcommand{\noi}{\noindent}

\newcommand{\be}{\begin{equation}}
\newcommand{\ee}{\end{equation}}
\newcommand{\bea}{\begin{eqnarray}}
\newcommand{\eea}{\end{eqnarray}}
\newcommand{\bes}{\begin{eqnarray*}}
\newcommand{\ees}{\end{eqnarray*}}

\newcommand{\bfi}{\begin{figure}}
\newcommand{\bfit}{\begin{figure}[t]}
\newcommand{\bfib}{\begin{figure}[b]}
\newcommand{\bfih}{\begin{figure}[h]}
\newcommand{\bfip}{\begin{figure}[p]}
\newcommand{\efi}{\end{figure}}

\newcommand{\bi}{\begin{itemize}}
\newcommand{\ei}{\end{itemize}}
\newcommand{\ben}{\begin{enumerate}}
\newcommand{\een}{\end{enumerate}}

\title{\LARGE \bf
Rate Distortion Function For a Class of Relative Entropy Sources
}

\author{Farzad Rezaei, Charalambos D. Charalambous and Photios A. Stavrou 
\thanks{The research leading to these results has received funding from the European
Community's Seventh Framework Programme (FP7/2007-2013) under grant agreement no. INFSO-ICT-223844.}
\thanks{F. Rezaei was at the University of Ottawa (\tt\small e-mail: frezaei@site.uottawa.ca).}
\thanks{C. D. Charalambous is with the Department of Electrical and Computer Engineering,
University of Cyprus, 75 Kallipoleos Avenue, P.O. Box 20537, 1678, Nicosia, CYPRUS (\tt\small email: chadcha@ucy.ac.cy). }%
\thanks{P. A. Stavrou is with the Department of Electrical and Computer Engineering,
University of Cyprus, 75 Kallipoleos Avenue, Nicosia, CYPRUS (\tt\small email: stavrou.fotios@ucy.ac.cy). }%
}

\begin{document}

\maketitle
\thispagestyle{empty}
\pagestyle{empty}

\begin{abstract}

This paper deals with rate distortion or source coding with fidelity criterion, in measure spaces, for a class of source distributions. The class of source distributions is described by a relative entropy constraint set between the true and a nominal distribution. The rate distortion problem for the class is thus formulated and solved using minimax strategies, which result in robust source coding with fidelity criterion. It is shown that minimax and maxmin strategies can be computed explicitly, and they are generalizations of the classical solution. Finally, for discrete memoryless uncertain sources, the rate distortion theorem is stated for the class omitting the derivations while the converse is derived.

\end{abstract}

\section{INTRODUCTION}

This paper is concerned with lossy data compression for a class of sources defined on the space of probability distributions on general alphabet spaces. In the classical rate distortion formulation with the fidelity decoding criterion, Shannon has shown that minimization of mutual information between finite alphabet source and reproduction sequences subject to fidelity criterion over the reproduction kernel has an operational meaning. Hence, it gives the minimum amount of information of representing a source symbol by a reproduction symbol with a pre-specified fidelity or distortion criterion.

The classical rate distortion function for
finite-alphabet and  continuous sources   has been studied
thoroughly in the literature \cite{berger71}, \cite{blahut72},
\cite{blahut87}, \cite{berger98} and \cite{csiszar74}. A survey of the
theory of rate distortion is given in  \cite{berger98}. The formulation of rate distortion function for abstract alphabets
is investigated by  Csisz\'ar in \cite{csiszar74}. Specifically, in
\cite{csiszar74} the question of existence of solution in Polish
spaces under some continuity assumptions on the distortion
function and compactness of the reproduction space, is established
under the topology of weak convergence. The formulation in
\cite{csiszar74} is based on two important assumptions, namely, 1)
compactness of the reproduction space, 2) absolute continuity of
all marginal distributions with respect to the optimal marginal
distribution. The compactness assumption is crucial in order to formulate the problem using countably
additive measures, and to show existence of the minimizing measure
using tightness arguments and Prohorov's theorem \cite{csiszar74}. Under these assumptions, the
optimal solution is derived and it is given by \bea q^*(x,dy)=
\frac{e^{s \rho(x,y)} \nu^*(dy)}{\int_{\hat{A}}e^{s \rho(x,z)}
\nu^*(dz)} \label{eq.i.1} \eea where $\rho$ is the distortion
function, $q^*$ is the optimal conditional distribution, $\nu^*$
is the optimal marginal distribution, $\hat{A}$ is the
reproduction space, and $s\in\Re$ is the Lagrange multiplier associated with the fidelity constraint.

 One of the fundamental issues for abstract alphabets, is
whether the nonlinear equation in (\ref{eq.i.1})
 has a solution. For the finite alphabet case, the existance of solution to ($\ref{eq.i.1}$) follows from the Blahut algorithm \cite{blahut72},
because in the limit the algorithm leads to an equation like
($\ref{eq.i.1}$). For general, abstract spaces (\ref{eq.i.1}) may not have solutions.
Clearly, if (\ref{eq.i.1}) does not have a solution, then the minimizing measure exists but
one cannot claim that it has the  form given by ($\ref{eq.i.1}$). Existence of a solution to the implicit nonlinear equation ($\ref{eq.i.1}$), is proved using Tihonov Fixed Point theorem, which holds for locally convex topological vector
spaces in \cite{farzad06}.

Source coding theorems with fidelity criteria  for abstract
sources are discussed in many papers. For separable metric spaces
results in this direction can be found in \cite{gray74}. This
result is applicable  to the set up considered in this paper.
Alternative approaches based on  Large Deviation techniques are
given in \cite{bucklew88}, while methods based on generalized AEP
(asymptotic equipartition property) are given in \cite{dembo2002}.
A source coding theorem for stationary source is presented in
\cite{gray77}. 

In \cite{sakrison69}, Sakrison extended the operational meaning of the rate distortion function to a class of sources. According to \cite{sakrison69}, when the class of sources is restricted to a compact class the rate distortion function of the class is precisely equal to the maximization over the class of the classical rate distortion function. Moreover, Sakrison's rate distortion function is calculated in \cite{wolf} for finite alphabet class of sources. Related subsequent work is also found in \cite{Munoz}, \cite{Neuh}.

This paper is concerned with the rate distortion or source coding problem with fidelity criterion on general abstract spaces, for a class of source distributions. The class of source distributions $\mu^{\prime}$ is modeled by a relative entropy $H(\cdot||\cdot)$, such that $H(\mu^{\prime}||\mu)\leq R$, $R>0$, where R is the distance from the so-called nominal source distribution $\mu$. The rate distortion for this class is formulated using minimax and maxmin strategies, with pay-off the mutual information between the source and reconstruction symbols, in which the minimum is with respect to the reconstruction conditional distribution (stochastic kernel), and the maximum is with respect to the source distribution $\mu^{\prime}$ which satisfies $H(\mu^{\prime}||\mu)\leq R$. 

Clearly, a class of source distributions defined by ${\cal M}_R(\mu)\tri \{ \mu^{\prime}\in {\cal M}_1(A); \quad H(\mu^{\prime}|| \mu) \leq R \}$, $R\geq0$, (${\cal M}_1(A)$ the set of probability distributions on A) is appealing since it is often used as a measure of distance between distributions, and $R_2\geq R_1$ implies ${\cal M}_{R_1}(\mu) \subseteq {\cal M}_{R_2}(\mu)$.

The objective is to compute both minimax and maxmin rate distortion functions for the class ${\cal M}_R(\mu)$ and to show operational meaning of the minimax rate distortion function by deriving a source coding theorem and its converse for this class of sources. The minimax and maxmin rate distortion functions are computed explicitly deriving expressions for the reproduction kernel which is a variant of (\ref{eq.i.1}). Moreover, from the solution it follows that both minimax and maxmin rate distortion yields the same answer. Due to space limitation the derivation of the source coding theorem is omitted and only the converse is presented.

\section{PROBLEM FORMULATION}

Assume $(A, {\cal A})$ and $(\hat{A}, \hat{{\cal A}})$ are two measurable spaces, where $A$ is the source space and $\hat{A}$ is the reproduction space. Assume $q:A \times \hat{{\cal A}} \rightarrow [0,1]$ is a mapping with the following two properties: \\
1) For every $x \in A$, the set function $q(x,.)$ is a probability measure on $\hat{{\cal A}}$. \\
2) For every $F \in \hat{{\cal A}}$, the function $q(.,F)$ is ${\cal A}$-measurable. \\
Mappings which satisfy 1) and 2) are called  stochastic kernels. Let ${\cal Q} (A,\hat{A})$ denote the class of all such stochastic kernels.\\
Given any measurable space $(\Sigma, {\bf {\Sigma}})$, let ${\cal M}_1(\Sigma)$ denote the space of probability measures on $\Sigma$.

Let $\mu \in {\cal{M}}_1(A)$ be the source probability. For a given pair $\{q \in {\cal Q}(A,\hat{A}),\mu \in {\cal M}_1(A)\}$ we can define three other probability measures as follows:\\
P1) The joint probability measure $P \in {\cal M}_1({ A} \times \hat{A})$ given by
\bes
P(G)=(\mu \otimes q) (G)= \int_{A} q(x,G_x) \mu(dx),  \quad \forall G \in {\cal A} \times \hat{{\cal A}}
\ees
where $G_x$ is the section of $G$ at point $x$, defined by $G_x \tri\{y \in \hat{A}: (x,y) \in G \}$ and $\otimes$ denotes convolution.\\
P2) The marginal probability measure $\nu \in {\cal M}_1(\hat{A})$  given by
\bes
\nu(F)\tri P(A \times F) &= \int_{A} q(x,(A \times F)_x) \mu(dx) \nonumber\\ &= \int_{A} q(x,F) \mu(dx), \quad \forall F \in \hat{{\cal A}}
\ees
P3) The product measure $\pi: {\cal A} \times \hat{{\cal A}} \rightarrow [0,1]$ of $\mu \in {\cal M}_1(A)$ and $\nu \in {\cal M}_1(\hat{A})$
\bes
\pi(G)=(\mu \times \nu)(G)= \int_A \nu(G_x) \mu(dx),   \quad \forall G \in {\cal A} \times \hat{{\cal A}}
\ees
Let $\rho: A \times \hat{A} \rightarrow [0, \infty)$ be a ${\cal A} \times \hat{{\cal A}}$-measurable function, and for each $D \in [0, \infty)$, define the set ${\cal Q}(D)$ as
\bes
\tilde{\cal Q}(D)= \{ q:A \times \hat{{\cal A}} \rightarrow [0,1]; \quad q(x,\hat{A})=1;\\\quad\int_{A} \int_{\hat{A}} \rho(x,y) q(x,dy) \mu(dx) \leq D \}
\ees
where each $q \in \tilde{\cal Q}(D)$ is ${\cal A}$-measurable for any $F \in \hat{{\cal A}}$ and $q(x,\hat{A})=1$ for any $x \in A$. For a given $P\in {\cal M}_1({ A} \times \hat{A})$ and $\mu \in {\cal M}_1(A)$ we assume that $\tilde{\cal Q}(D)$ is non empty.\\
Given a fixed source measure $\mu \in {\cal M}_1(A)$ the rate distortion function is defined as follows
\bea
R(D)= \inf_{q \in \tilde{\cal Q}(D)} H(P || \pi) \label{f1}=\inf_{q \in \tilde{\cal Q}(D)}I(\mu ; q)
\eea
where $H(P || \pi)$ is the relative entropy between $P$ and $\pi$ and is denoted by $I(\mu ; q)$. More explicitly, by using
\bes
&&P(dx \times dy) = \mu(dx) \otimes q(x,dy) \\
&&\pi(dx \times dy)= \mu(dx) \otimes \nu(dy)
\ees
$R(D)$ is given by
\bea
&&R(D)= \inf_{q \in \tilde{\cal Q}(D)} I(\mu;q) \nonumber\\&&=\inf_{q \in \tilde{\cal Q}(D)} \int_{A} \int_{\hat{A}} \log \Big( \frac{q(x,dy)}{\nu(dy)}\Big) q(x,dy) \mu(dx) \label{f2}
\eea
where for every $q$ in $\tilde{\cal Q}(D)$ we have
\bes
&&\int_{A \times \hat{A}} \rho(x,y) P(dx,dy) \leq D \quad \mbox{ or}\\
&&\int_{A} \int_{\hat{A}} \rho(x,y) q(x,dy) \mu(dx) \leq D
\ees


\section{RATE DISTORTION FOR A FIXED SOURCE}

Throughout the rest of the paper we assume that both $A$ and $\hat{A}$ are polish spaces (complete, separable metric spaces) and so normal topological spaces. The following theorem found in \cite{farzad06} is a generalization of \cite{csiszar74} relaxing the assumptions of compactness and absolute continuity while identifying appropriate function spaces in which the solution is sought.
\begin{theorem} \label{th1}
\noi Let $A$, $\hat{A}$ be two polish spaces at $\rho: A \times \hat{A} \rightarrow [0, \infty]$ a measurable, nonnegative extended real-valued function, continuous in the second argument, and $\mu \in {\cal M}_1(\Sigma)$ be fixed. Then\\
1) \bes R(D)=\inf_ {q \in \tilde{\cal Q}(D)}I(\mu ; q) \ees has a solution.\\
2) Suppose the set $F \tri \{ (x,y) \in A \times \hat{A} ;\rho (x,y)<D \}$ is non-empty. Then the constraint problem $R(D)$ is equivalent to the unconstraint problem.
\bes
&&R(D)= \max_{s \leq 0} \inf_{q \in \tilde{\cal Q}(D)} \{I(\mu ; q) \\&&-s \Big( \int_{A} \int_{\hat{A}} \rho(x,y) q(x,dy) \mu(dx) - D \Big) \Bigg\}
\ees
Further, the infimum occurs on the boundary of the set $\tilde{\cal Q}(D)$ and the infimum is attained at
\bes
q^*(x,F)=\frac{\int_{F} e^{s \rho(x,y)} \nu^*(dy)}{\int_{\hat{A}} e^{s \rho(x,y)} \nu^*(dy)},&& s\leq 0
\ees
The maximization over $s\leq 0$ denoted by $s^*$ is found from the constraint which is satisfied with equality.
The corresponding rate distortion function has the following form
\bes
R(D)= s^*D - \int_{A}\log \Big( \int_{\hat{A}} e^{s^* \rho(x,y)} \nu^*(dy)\Big) \mu(dx)
\ees
\end{theorem}
Note that the solution presented in Theorem $\ref{th1}$ is one form of the rate distortion solution. Alternative expressions are found in \cite{berger71}. The main objective of this paper is to extend the results of Theorem $\ref{th1}$ to a class of sources described by a relative entropy constraint set, and to show source coding theorem and its converse.

\section{RATE DISTORTION FOR A CLASS OF SOURCES}

Let $\mu \in {\cal M}_1(A)$ denote the nominal (fixed) probability measure which is not the true source probability measure. Further, assume the true source probability measure belongs to the following relative entropy constraint set
\bes
{\cal M}_R(\mu)\tri \Big\{ \mu^{\prime}\in {\cal M}_1(A); \quad H(\mu^{\prime}|| \mu) \leq R \Big\}
\ees
where $R\geq0$ is given and fixed, in $[0, \infty)$. Clearly the larger $R$ is the larger the class of distributions allowed in the set. In the absense of uncertainty, the set ${\cal M}_R(\mu)$ reduces to the singleton $\{\mu\}$. For a given $q \in {\cal Q}(D)$ and a given $\mu \in {\cal M}_1(A)$ let $P^{\prime}\in {\cal M}_1(A \times \hat{A})$ denote the joint probability measure defined by
\bes
P^{\prime}(G)=(\mu^{\prime} \otimes q) (G)= \int_{A} q(x,G_x) \mu^{\prime}(dx),\quad \forall G \in {\cal A} \times \hat{{\cal A}}
\ees
Also define the marginal probability measure $\nu^{\prime} \in {\cal M}(\hat {A})$ by
\bes
\nu^{\prime}(F)=P^{\prime}(A \times F) &&= \int_{A} q(x,(A \times F)_x) \mu^{\prime}(dx) \nonumber\\&&= \int_{A} q(x,F) \mu^{\prime}(dx), \quad \forall F \in \hat{{\cal A}}
\ees
Denote the product of $\mu^{\prime}\in {\cal M}_1(A)$ and $\nu^{\prime}\in {\cal M}_1(\hat {A})$ by $\pi^{\prime}$, defined by
\bes
\pi^{\prime}(G)=(\mu^{\prime} \times \nu)(G)= \int_A \nu(G_x) \mu^{\prime}(dx),   \quad  \forall G \in {\cal A} \times \hat{{\cal A}}
\ees
Let $\rho: A \times \hat{A} \rightarrow [0, \infty)$ be a ${\cal A} \times \hat{{\cal A}}$-measurable function, and for each $D \in [0, \infty)$, define the set ${\cal Q}(D)$ as
\bes
{\cal Q}(D)= \{ q:A \times \hat{{\cal A}} \rightarrow [0,1]; \quad q(x,\hat{A})=1 ; \\\quad \int_{A} \int_{\hat{A}} \rho(x,y) q(x,dy) \mu^{\prime}(dx) \leq D ,\quad \forall \mu^{\prime} \in {\cal M}_R(\mu)\}
\ees
where each $q \in {\cal Q}(D)$ is ${\cal A}$-measurable for any $F \in \hat{{\cal A}}$ and $q(x,\hat{A})=1$ for any $x \in A$, and $D \in [0, \infty)$ and $\rho: A \times \hat{A} \rightarrow [0,\infty)$, is a non-negative measurable function with respect to the measurable space ${\cal A} \times \hat{\cal A}$.\\
Given the class ${\cal M}_R(\mu)$ of uncertain source probability measures the Rate Distortion for the class of ${\cal M}_R(\mu)$ is defined by
\bea
&&R_{+}(D)\tri \inf_{q \in {\cal Q}(D)} \sup_{\mu^{\prime} \in {\cal M}_R(\mu)} I(\mu^{\prime};q) \label{f30}
\eea
Note that in the minimax formulation of Rate Distortion the uncertainty $\mu^{\prime} \in {\cal M}_R(\mu)$ tries to maximize the rate of reconstructing the source while the designer $q \in {\cal Q}(D)$ tries to minimize the rate. Thus, $R_{+}(D)$ is the rate distortion of the class ${\cal M}_R(\mu)$.\\
An alternative formulation is to consider the maxmin Rate Distortion
 \bea
 &&R_{-}(D)\tri \sup_{\mu^{\prime} \in {\cal M}_R(\mu)} \inf_{q \in {\cal Q}(D)} I(\mu^{\prime};q) \label{f31}
 \eea
It can be shown that $R_{+}(D)\geq R_{-}(D)$ while equality holds if the minisup Theorem $\ref{th7}$ (see Appendix) holds. It can be easily shown that by formulating $R_{+}(D)$, $R_{-}(D)$ using countably additive probability measures and weak convergence as in \cite{csiszar74} or regular bounded finitely additive probability measures and weak$^*$ convergence as in \cite{farzad06}, that the conditions of minisup theorem, Theorem $\ref{th7}$ (Appendix) hold. Hence, $R_{+}(D)=R_{-}(D)$. Nevertheless, in the next two Theorems we find the minimax and maxmin strategies and then verify using these strategies that $R_{+}(D)=R_{-}(D)$. Once these strategies are obtained and $R_{+}(D)=R_{-}(D)$ is established, then the solution of $R_{-}(D)$ is used to prove the coding theorem.

Below we provide the solutions to $R_{+}(D)$ and $R_{-}(D)$.

\begin{theorem} \label{th5}
Suppose $e^{\frac{\ell}{\lambda}} \in L_1(\mu)$ and $\ell e^{\frac{\ell}{\lambda}} \in L_1(\mu)$, where $\ell(x)=-\log \Big( \int_{\hat{A}} e^{s \rho(x,y)} \nu^*(dy)\Big),\quad \lambda\geq 0$. Then the infimum and supremum of $R_{-}(D)$ are attained by the following distributions:
\bes
&&\mu^*(dx)= \frac{ \Big( \frac{1}{\int_{\hat{A}} e^{s \rho(x,y)} \nu(dy)}\Big)^{\lambda}\mu(dx)}{\int_{A} \Big( \frac{1}{\int_{\hat{A}} e^{s \rho(u,y)} \nu^*(dy)}\Big)^{\lambda}\mu(du)},\quad \lambda\geq 0 \\
&&q^*(x,dy)= \frac{e^{s \rho(x,y)} \nu^*(dy)}{\int_{\hat{A}} e^{s \rho(x,z)} \nu^*(dz)},\quad s\leq 0
\ees
where $s\leq 0$, $\lambda \geq 0$ are found from the constraints.\\
The rate distribution $R_{-}(D)$ is given by
\bes
R_{-}(D)= sD + \lambda R + \lambda \log \int_{A} \Big( \int_{\hat{A}} e^{s \rho(x,y)} \nu^*(dy) \Big)^{\frac{-1}{\lambda}} \mu(dx)
\ees
\end{theorem}
\noi {\bf Proof.} See Appendix.

The following Lemma is needed to be able to apply Theorem $\ref{th8}$ to find the solution of $R_{+}(D)$.

\begin{lemma} \label{lem1}
Assume $e^{s \rho} \in L_1(\nu)$, $s\leq 0$ then $\ell$ define by
\bes
\ell(x) \tri \int_{\hat{A}} \log \Bigg( e^{-s \rho(x,y)} \frac{q(x,dy)}{\nu^*(dy)}\Bigg) q(x,dy),\quad s\leq 0
\ees
is bounded below.
\end{lemma}
\noi{\bf Proof.} Omitted.

 Using Lemma $\ref{lem1}$ and applying Theorem $\ref{th8}$ (see Appendix) similar to Theorem $\ref{th5}$, we deduce the solution of $R_{+}(D)$.
\begin{theorem} \label{th6}
Suppose $e^{s \rho} \in L_1(\nu)$, $s\leq 0$ then the supremum and infimum of $R_{+}(D)$ are attained by the following distributions:
\bes
&&d\mu^*=  \frac{e^{\frac{\ell}{\lambda}} d\mu}{\int_{A}e^{\frac{\ell}{\lambda}} d\mu},\quad \lambda\geq 0 \\
&&q^*(x,dy)= \frac{e^{s \rho(x,y)} \nu^*(dy)}{\int_{\hat{A}} e^{s \rho(x,z)} \nu^*(dz)},\quad s\leq 0
\ees
where $\ell$ is defined by
\bes
\ell(x)\tri \int_{\hat{A}} \log \Bigg( e^{-s \rho(x,y)} \frac{q^*(x,dy)}{\nu^*(dy)}\Bigg) q^*(x,dy)
\ees
and $s\leq 0$, $\lambda \geq 0$ are found from the constraints.\\
The rate distortion $R_{+}(D)$ is given by
\bes
R_{+}(D)= sD + \lambda R + \lambda \log \int_{A} \Big( \int_{\hat{A}} e^{s \rho(x,y)} \nu^*(dy) \Big)^{\frac{-1}{\lambda}} \mu(dx)
\ees
\end{theorem}
\noi {\bf Proof.} Follows as in Theorem $\ref{th5}$.

\begin{lemma} \label{lem2}
For any distribution $\mu^{\prime}$ in the set ${\cal M}_R(\mu)$, we have
\bes
H(\mu^{\prime} || \mu^*) \leq R^*
\ees
where $\mu^*$ is the source distribution found in Theorem $\ref{th5}$, with $\nu$ replaced by $\nu^*$ and $R^*$ is given by
\bes
&&R^*=\log \Bigg( \int_{A} \Big( \int_{\hat{A}} e^{s \rho(x,y)} \nu^*(dy) \Big)^{-\lambda} \mu(dx) \Bigg)+ R \\&& + \lambda \Bigg( \alpha R + \alpha \log \Big( \int_{A} \Big( \int_{\hat{A}} e^{s \rho(x,y)} \nu^*(dy) \Big)^{\frac{1}{\alpha}} \mu(dx) \Big) \Bigg)
\ees
where $\lambda, \alpha \in \Re$ are constants.
\end{lemma}
\noi{\bf Proof.} See Appendix.

Next we state the rate distortion theorem for uncertain discrete memoryless sources with distributions in ${\cal M}_R(\mu)$. The derivation follows the same steps as in \cite{berger71}.

\begin{theorem} {\bf Robust source coding theorem}
Let the set of discrete memoryless sources $\{X, \mu^{\prime} \}$ with $H(\mu^{\prime}|| \mu) \leq R$, and single letter fidelity criterion be given. Let $R^*(D)$ denote the robust rate distortion function defined in Theorem~\ref{th5} or Theorem~\ref{th6}. Then given any $\epsilon > 0$ and any $D \geq 0$, an integer $n$ and a source code with block length $n$, and rate ${\cal{R}} < R^*(D)+ \epsilon$ exists, such that for any distribution from the set ${\cal M}_R(\mu)$, the code is $D+ \epsilon$-admissible.
\end{theorem}
\noi {\bf Proof.}
The basic idea for the proof is to construct the code based on $q^*$ and $\nu^*$, found in the robust rate distortion formulation ( see Theorem~\ref{th5}). However, due to the uncertainty in the source distribution, all the averages are taken with respect to $\mu^{\prime}$ and $q^*$. Then   these averages are related to averages which appear in the rate distortion theorem for $\mu^*$ and $q^*$, which are the solutions to the robust rate distortion problem. The proof is based on a random coding argument.

\begin{theorem}{\bf Converse to the robust source coding theorem}
Every code which is $D$-admissible for the whole class of source distributions ${\cal M}_R(\mu)$, has rate greater than $R^*(D)$, i.e.,
\bes
\frac{1}{n} \log K(n,D) \geq R^*(D)
\ees
where $K(n,D)$ is the number of $D$-admissible codewords of length $n$, in the code.
\end{theorem}
\noi {\bf Proof.}
Suppose code ${\cal C}$ with rate ${\cal{R}} = \frac{1}{n} \log K(n,D)$ is $D$-admissible for the whole class of source distributions ${\cal M}_R(\mu)$, Then by the converse source coding theorem for a fixed source distribution $\mu^{\prime}$ from the set, we have
\bea
\frac{1}{n} \log K(n,D) \geq R_{\mu^{\prime}}(D) \label{f51}
\eea
where $R_{\mu^{\prime}}(D)$ is the rate distortion function for the source with distribution $\mu^{\prime}$. Since our code is $D$-admissible for any $\mu^{\prime} \in {\cal M}_R(\mu)$, then ($\ref{f51}$) must hold for all $\mu^{\prime} \in {\cal M}_R(\mu)$.
\bes
\frac{1}{n} \log K(n,D) \geq R_{\mu^{\prime}}(D), \quad \forall \mu^{\prime} \in {\cal M}_R(\mu)
\ees
Taking supremum of both sides with respect to $\mu^{\prime}$ leads to
\bes
\frac{1}{n} \log K(n,D) \geq {\sup_{\mu^{\prime} \in {\cal M}_R(\mu)}} R_{\mu^{\prime}}(D)
\ees
By Theorem~\ref{th5}, we have
$\frac{1}{n} \log K(n,D) \geq  R^*(D)$.

\section{CONCLUSION}


The problem of rate distortion is extended to the case of uncertain sources, in which the uncertainty description about the true source distribution is described by a relative entropy constraint set between the true and a nominal distribution. The rate distortion problem is thus formulated and solved using minimax strategies, which results in robust source coding with fidelity criterion. The solution is found for both minimax and maxmin strategies. Finally, for discrete memoryless uncertain sources, the rate distortion theorem is stated and its converse is proved.

%

%

\section{APPENDIX}

The next minisup Theorem states necessary conditions for $R_{+}(D)=R_{-}(D)$.
\begin{theorem} \label{th7}{\bf Minisup Theorem} \cite{Fan53}
Let $f(x,y)$ be defined for $x \in \cal{X}$,  $y \in \cal{Y}$, where $\cal{X}$ and $\cal{Y}$ are convex subsets of topological vector spaces and $\cal{X}$ is compact, $f(x,y)$ be convex and lower semicontinuous in $x \in \cal{X}$ for each $y \in \cal{Y}$ and concave in $y \in \cal{Y}$ for each $x \in \cal{X}$.\\
Then there exists an $x^* \in \cal{X}$ such that
\bes
\sup_{y \in \cal{Y}} \min_{x \in \cal{X}} f(x,y)=\sup_{y \in \cal{Y}} f(x^*,y)=\min_{x \in \cal{X}} \sup_{y \in \cal{Y}} f(x,y)
\ees
\end{theorem}

The next theorem gives the duality between relative entropy and free energy.
\begin{theorem} \label{th8}
\cite{dupuis-ellis97} For every $\ell:\Sigma \rightarrow \Re$ measurable function bounded below and $\mu \in {\cal M}_1(\Sigma)$. Then
\bes
&&\sup_{\{\nu \in {\cal M}_1(\Sigma); H(\nu||\mu)<\infty\}}\big\{\int_{\Sigma}\ell(x)\nu(dx)-H(\nu||\mu)\big\}\\&&=\log\int_{\Sigma}e^{\ell(x)} \mu(dx)
\ees
Moreover, if $\ell e^{\ell} \in {\cal M}_1(\mu)$ then the supremum is attained by the tilted $\nu \in {\cal M}_1(\Sigma)$ given by
\bes
\nu^*(dx)=\frac{e^{\ell(x)}\mu(dx)}{\int e^{\ell(x)}\mu(dx)}
\ees
\end{theorem}

\noi {\bf Proof of Theorem~\ref{th5}.}\\
 By Theorem~\ref{th1}, we already know that
\bes
&&R_{\mu^{\prime}}(D)\tri \inf_{q \in {\cal Q}(D)} I(\mu^{\prime}; q)\\&&= {\max_{s \leq 0}}\big\{sD - \int_{A}\log \Big( \int_{\hat{A}} e^{s \rho(x,y)} \nu^*(dy) \Big) \mu^{\prime}(dx)\big\}
\ees
Now consider the function $\ell(x)\tri -\log \Big( \int_{\hat{A}} e^{s \rho(x,y)} \nu^*(dy) \Big)$. From previous results, we know that $s \leq 0$, hence $\ell(x) \geq 0$, and therefore $\ell(x)$ is a bounded-below measurable function defined on the measurable space $(A, {\cal A})$. Now we solve the problem $R_{-}(D)$, using Lagrange multilpiers.
\bes
&&R_{-}(D)=\sup_{\mu^{\prime} \in {\cal M}_R(\mu)} R_{\mu^{\prime}}(D)\\&&=\min_{\lambda\geq 0}\sup_{\mu^{\prime} \in {\cal M}_R(\mu)}\max_{s\leq 0} \Big( R_{\mu^{\prime}}(D) - \lambda (H(\mu^{\prime}|| \mu)-R)\Big)
\\&&= \min_{\lambda\geq 0}\sup_{\mu^{\prime} \in {\cal M}_R(\mu)}\max_{s\leq 0}\Big\{\Big( \int_{A} \ell d\mu^{\prime}- \lambda (H(\mu^{\prime}|| \mu)-R) \Big)\\&&+ sD \Big\}
\ees
Now we use the duality between relative entropy and free-energy as explained in Theorem~\ref{th8} to find the above supremum.
If $e^{\frac{\ell}{\lambda}} \in L_1(\mu)$ and $\ell e^{\frac{\ell}{\lambda}} \in L_1(\mu)$, $\lambda\geq 0$ then the supremum is attained for $\mu^*$ given by
\bes
\mu^*(dx)&= \frac{e^{\frac{\ell(x)}{\lambda}} \mu(dx)}{\int_{A} e^{\frac{\ell(x)}{\lambda}} \mu(dx)}&= \frac{ \Big( \frac{1}{\int_{\hat{A}} e^{s \rho(x,y)} \nu^*(dy)}\Big)^{\lambda}\mu(dx)}{\int_{A} \Big( \frac{1}{\int_{\hat{A}} e^{s \rho(u,y)} \nu^*(dy)}\Big)^{\lambda}\mu(du)}
\ees
Hence
\bea
&&R_{-}(D)=\sup_{\mu^{\prime} \in {\cal M}_R(\mu)} R_{\mu^{\prime}}(D)\nonumber\\&&=\min_{\lambda\geq 0}\max_{s\leq 0}\Big\{ sD + \lambda R + \lambda \log \Big( \int_{A} e^{\frac{\ell}{\lambda}} d\mu\Big)\Big\} \label{f32}
\eea
where min over $\lambda\geq 0$ denoted by $\lambda^*$ is chosen such that $H(\mu^*||\mu)\mid_{\lambda=\lambda^*}=R$, and max over $s\leq 0$ denoted by $s^*$ is chosen such that the distortion constraint holds with equality. Now ($\ref{f32}$), can be written as
\bes
R_{-}(D)=\sup_{\mu^{\prime} \in {\cal M}_R} R_{\mu^{\prime}}(D)= s^*D + \lambda^*R +\\+ \lambda^*\log \int_{A} \Big( \int_{\hat{A}} e^{s^*\rho(x,y)} \nu^*(dy) \Big)^{\frac{-1}{\lambda^*}} \mu(dx)
\ees
Also by Theorem~\ref{th1}, the reproduction kernel $q^*$ for the rate distortion problem defined for the source $\mu^*$ is given by
\bes
q^*(x,dy)= \frac{e^{s^*\rho(x,y)} \nu^*(dy)}{\int_{\hat{A}} e^{s^* \rho(x,z)} \nu(dz)}
\ees


\noi {\bf Proof of Lemma~\ref{lem2}.} \\
\bes
H(\mu^{\prime} || \mu^*) &=& \int_{A} \log \Big( \frac{d\mu^{\prime}}{d\mu^*}\Big) d\mu^{\prime}  \\
				 &=& \int_{A} \log \Big( \frac{d\mu^{\prime}}{d\mu}\Big) d\mu^{\prime} + \int_{A} \log \Big( \frac{d\mu}{d\mu^*}\Big) d\mu^{\prime}
\ees
Now $\frac{d\mu}{d\mu^*}$ can found from Theorem $\ref{th5}$, so we have
\bea
&H(\mu^{\prime} || \mu^*) = H(\mu^{\prime} || \mu) \nonumber\\&+ \int_{A} \log \Bigg( \frac{\int_{A}\Big( \int_{\hat{A}} e^{s\rho(z,y)} \nu^*(dy) \Big)^{-\lambda} \mu(dz)  }{\Big( \int_{\hat{A}} e^{s\rho(x,y)} \nu^*(dy) \Big)^{-\lambda}} \Bigg) \mu^{\prime}(dx) \nonumber \\
                  &= H(\mu^{\prime} || \mu) + \log \Bigg( \int_{A}\Big( \int_{\hat{A}} e^{s\rho(z,y)} \nu^*(dy) \Big)^{-\lambda} \mu(dz)\Bigg)\nonumber \\&+ \lambda \int_{A} \log \Big( \int_{\hat{A}} e^{s \rho(x,y)} \nu^*(dy) \Big) \mu^{\prime}(dx)  \label{f34}
\eea
In ($\ref{f34}$), we can find the supremum of the right-hand side term
\bes
J= \sup_{\mu^{\prime} \in {\cal M}_R} \int_{A} \log \Big( \int_{\hat{A}} e^{s \rho(x,y)} \nu^*(dy) \Big) \mu^{\prime}(dx)
\ees
The function inside the first integral is measurable and bounded below by zero. Let
\bes
\ell(x)=\log \Big( \int_{\hat{A}} e^{s \rho(x,y)} \nu^*(dy) \Big)
\ees
Then
\bea
J= \sup_{\mu^{\prime} \in {\cal M}_R} \int_{A} \ell d\mu^{\prime} = \alpha R + \alpha \log \Big( \int_{A} e^{\frac{\ell}{\alpha}} d\mu \Big) \label{f35}
\eea
where $\alpha$ is chosen in a way that $H(\mu^{\prime,*}|| \mu)=R$ for the following measure
\bes
d\mu^{\prime,*}= \frac{e^{\frac{\ell}{\alpha}}d\mu}{\int_{A} e^{\frac{\ell}{\alpha}} d\mu }
\ees
Combine ($\ref{f34}$) and ($\ref{f35}$) to get
\bea
H(\mu^{\prime} || \mu^*) \leq \log \Bigg( \int_{A} \Big( \int_{\hat{A}} e^{s \rho(x,y)} \nu^*(dy) \Big)^{-\lambda} \Bigg) + R\nonumber\\
+ \lambda \Bigg( \alpha R + \alpha \log \Big( \int_{A} \Big( \int_{\hat{A}} e^{s \rho(x,y)} \nu^*(dy) \Big)^{\frac{1}{\alpha}} \mu(dx) \Big) \Bigg) \label{f36}
\eea
Also the supremum in ($\ref{f35}$) is achieved for $\mu^{\prime,*}$, and for this measure we have $H(\mu^{\prime,*} || \mu)=R$. Therefore the right hand side of ($\ref{f36}$) is achieved for $\mu^{\prime,*}$,  ad finally
\bes
H(\mu^{\prime,*}||\mu^*) = R^*
\ees
where
\bes
&R^*=\log \Bigg( \int_{A} \Big( \int_{\hat{A}} e^{s \rho(x,y)} \nu^*(dy) \Big)^{-\lambda} \Bigg) \\&+R+ \lambda \Bigg( \alpha R + \alpha \log \Big( \int_{A} \Big( \int_{\hat{A}} e^{s \rho(x,y)} \nu^*(dy) \Big)^{\frac{1}{\alpha}} \mu(dx) \Big) \Bigg)
\ees

\end{document}